\documentclass[runningheads]{llncs}
\usepackage[T1]{fontenc}
\usepackage{graphicx}
\usepackage{multicol}
\usepackage{multirow}
\usepackage{makecell} 
\usepackage{booktabs,lipsum, longtable}
\usepackage{algpseudocode, algorithm, xcolor}
\usepackage{amsmath}
\usepackage{cite}
\begin{document}
 \title{NeuralMultiling: A Novel Neural Architecture Search for Smartphone based Multilingual Speaker Verification}
%
%
\author{Aravinda Reddy PN\inst{1}\orcidID{0000-0002-1342-924x} \and
Raghavendra Ramachandra\inst{2}\orcidID{0000-0003-0484-3956} \and K.Sreenivasa Rao \inst{1}\orcidID{0000--0001-6112-6887} \and Pabitra Mitra\inst{1}\orcidID{0000--0002-1908-9813}   }
\authorrunning{Aravinda et al.}
%
\institute{Indian Institute of Technology Kharagpur, India\\
\email{aravindareddy.27@iitkgp.ac.in}\\
\and
Norwegian University of Science and Technology (NTNU), Norway.}
\maketitle              
\begin{abstract}
Multilingual speaker verification introduces the challenge of verifying a speaker in  multiple languages. Existing systems were built using i-vector/x-vector approaches along with Bi-LSTMs, which were trained to discriminate speakers, irrespective of the language. Instead of exploring the design space manually, we propose a neural architecture search for multilingual speaker verification suitable for mobile devices, called \textbf{NeuralMultiling}. First, our algorithm searches for an optimal operational combination of neural cells with different architectures for normal cells and reduction cells and then derives a CNN model by stacking neural cells. Using the derived architecture, we performed two different studies:1) language agnostic condition and 2) interoperability between languages and devices on the publicly available Multilingual Audio-Visual Smartphone (MAVS) dataset. The experimental results suggest that the derived architecture significantly outperforms the existing Autospeech method by a 5-6\% reduction in the Equal Error Rate (EER) with fewer model parameters.

\end{abstract}

\section{Introduction}
\label{sec:intro}

 Biometric-based secure verification is widely deployed in many applications, such as door locks, security devices, home automation, IoT, smart speakers, game consoles, border control, smartphone unlocking, banking, and financial transactions. Over the years, the evolution of smartphones has enabled the biometric-based secure verification of several financial applications, including banking transactions.  Biometric verification of smartphones can be achieved using physiological and behavioral biometrics. The most commonly used biometric characteristics in smartphone verification include faces \cite{rattani2018survey}, irises or eyes \cite{8698587}, fingerphotos \cite{ramachandra2019smartphone} and voice \cite{mandalapu2021multilingual}. Each biometric characteristic has its own advantages and disadvantages in terms of usability, accuracy, and user experience.

Voice-based biometric verification is widely employed in various smartphone applications including  banking \cite{Voice:2023}. The main advantages of using voice biometrics in smartphone applications are accuracy, scalability, and usability. Conventional voice biometric systems enrol speakers in one language by using short sentences. During the verification, the speaker will utter the same sentence (in the case of text-dependent) or different sentences (in the case of text independence) in the same language used during the enrolment, which will be compared with the enrolled sample to make the verification decision. However, the use of the same language limits both service providers (or vendors) and users, in terms of scalability and usability. Because users can speak more than one language at a time, it is more convenient for speakers to use multilingual verification than a single language. From a vendor’s perspective, it is important to build language-independent models that can achieve scalability. These factors motivated multilingual speaker verification, which allows the user to enrol in one language and verify it with another language. Therefore, multilingual speaker verification aims to verify speaker identity based on speech utterances from one or more languages and ensure that the voice-based security system is robust and generalizable to various applications \cite{mandalapu2021multilingual}.

\begin{figure}[!t]
  \centering
  \includegraphics[width=1\linewidth]{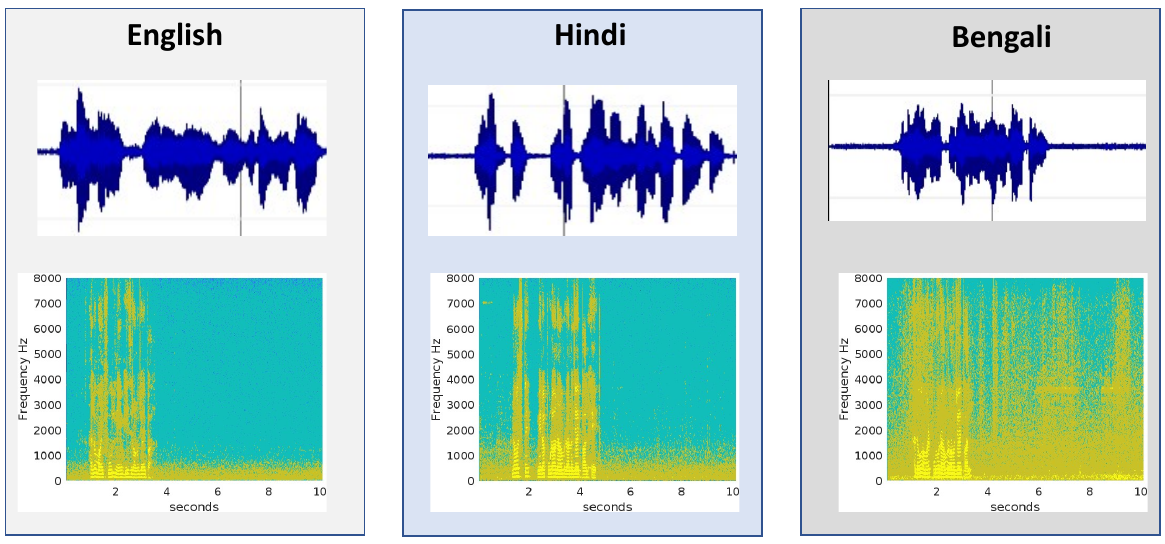}
  \caption{Illustration of speech signal and corresponding spectrogram of the different languages uttered by the same subject}
  \label{fig:intro}
\end{figure}

Figure \ref{fig:intro} illustrates example time signals and corresponding spectrograms for different languages uttered by the same subject. The different characteristics of the language, especially the sequence of phonemes, and  the language-specific spectral characteristics of utterances from the same subject introduce challenges for reliable multilingual verification. However, as the same speaker is speaking multiple languages, even though the sequence of phonemes may differ in languages, the individual phoneme characteristics may remain somewhat the same across languages because of the geometry offered by the same vocal tract, which has motivated researchers to develop multilingual speaker verifications.  Earlier studies have explored multilingual speaker verification systems by extracting i-vectors or x-vectors from speech utterances \cite{dehak2010front, snyder2018x}, which are trained to discriminate speakers and map utterances to fixed dimensional embeddings. Recently, the 2D CNN model \cite{li2017cross} was trained on English data and evaluated in the Chinese and Uyghur languages. It significantly outperforms the baseline i-vector model by a large margin \cite{li2017cross}.

The deployment of a multilingual speaker-verification model for smartphones is challenging because of the requirement for a lightweight text-independent model that can be generalized across different languages.  In this study, we propose a novel CNN architecture for multilingual speaker verification based on  Neural Architecture Search (NAS) methods \cite{liu2018darts} to derive the best CNN architecture for text-independent multilingual speaker verification.
The proposed method has two novel features 1) An automatic network search that can result in an optimized network architecture for a multilingual speaker verification model. 2) A different architecture for normal and reduction cells to achieve reliable multilingual speaker verification with a lightweight model.

The main contributions of this study are as follows.
\begin{enumerate}
 \item Novel method for multilingual speaker verification using differentiable neural architecture search to achieve the optimized lightweight model. 
 \item The proposed method is initialized to have different architecture for normal and reduction cells to better quantify the speaker characteristics. 
 \item Extensive experiments are presented on the publicly available MAVS dataset with 37,800 utterances representing three different languages. The MAVS dataset was collected in three different sessions using five different smartphones from 103 subjects with unique data. 
 \item The performance of the proposed method is compared with the Autospeech \cite{ding2020autospeech}, which derives the architecture in automated way.
\end{enumerate}

The rest of the paper is organized as follows:  Section \ref{sec:RelatedWork} discusses the related work on multilingual speaker verification, Section \ref{Method} presents the proposed method, Section \ref{sec:exp} discusses the experimental protocols, architecture search, and quantitative results, and Section \ref{concl} concludes the paper.

\vspace{-0.4cm}
\section{Related work}
\label{sec:RelatedWork}

Multilingual speaker verification has attracted significant interest from researchers in the recent decades. Early work began with the introduction of the first Spanish corpus named AHUMADA \cite{ortega2000ahumada} by NIST \cite{greenberg2020two}.  Joint Factor Analysis (JFA) was adopted by \cite{lu2009effect}, in which language factors were captured in training and testing utterances. The evaluation results showed a significant improvement in the performance of the non-English trials. 



The NIST Speaker Recognition Challenge in 2016 revealed the importance of score normalization for mismatched data conditions. Therefore, \cite{matejka2017analysis} compared several normalization techniques, as well as different cohorts, and analyzed the nature of the files selected for the cohort in adaptive score normalization. Unsupervised speaker verification was conducted using adversarial training \cite{xia2019cross}. For short utterances, hard prototype mining as a computationally efficient hard negative mining strategy to fine-tune the x-vectors was adopted by \cite{thienpondt2020cross}. A large-scale study of 46 languages was conducted by \cite{chojnacka2021speakerstew}, in which a hybrid novel triage mechanism was introduced for both text-dependent and text-independent methods. Lately \cite{nam2022disentangled} proposed disentangled representation learning to disentangle speaker module and language module. Both modules have a speaker feature extractor, embedding layer, and classifier to achieve reliable speaker verification.

It is worth noting that all existing studies are mainly focused on a non-smartphone environment, where the  requirement of lightweight models is of paramount importance.  Recently, \cite{mandalapu2021multilingual} benchmarked a smartphone-based SWAN dataset consisting of four different languages by performing a cross-lingual speaker verification using the x-vector method. However, the development of lightweight models is important, particularly in smartphone environments. The first neural architecture search-based speaker recognition autospeech was recently proposed by \cite{ding2020autospeech}. The experimental results indicate a lightweight model with robust performance in English, which  motivated us to propose a neural search method for multilingual speaker verification. \textit{We hypothesize that ameliorating network architecture design matters for deriving a lightweight model for multilingual speaker verification for mobile devices}. Therefore, we considered Autospeech \cite{ding2020autospeech} as a baseline model and proposed a modified architecture that considers normal cells and reduction cells to have different architectures to increase the search space within the specified space, by which better speaker characteristics are captured.

\begin{figure}[!ht]
  \centering
  \includegraphics[width=1\linewidth]{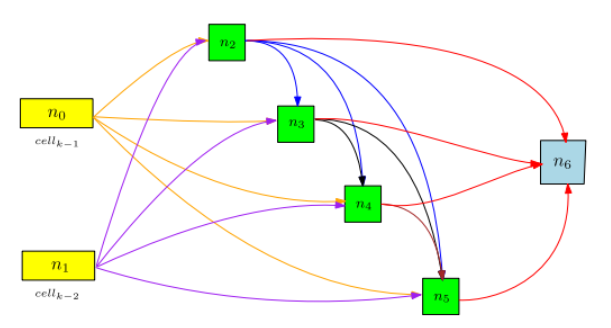}
  \caption{Depiction of a neural cell. The transitional nodes($x_2$ to $x_5$) are thickly connected during the search process. Only two operations with the highest softmax probabilities are retained during architecture derivation for the transitional nodes.}
  
  \label{fig:sample_cell}
\end{figure}

\begin{figure*}[!t]
    \centering
    \includegraphics[width=1 \linewidth]{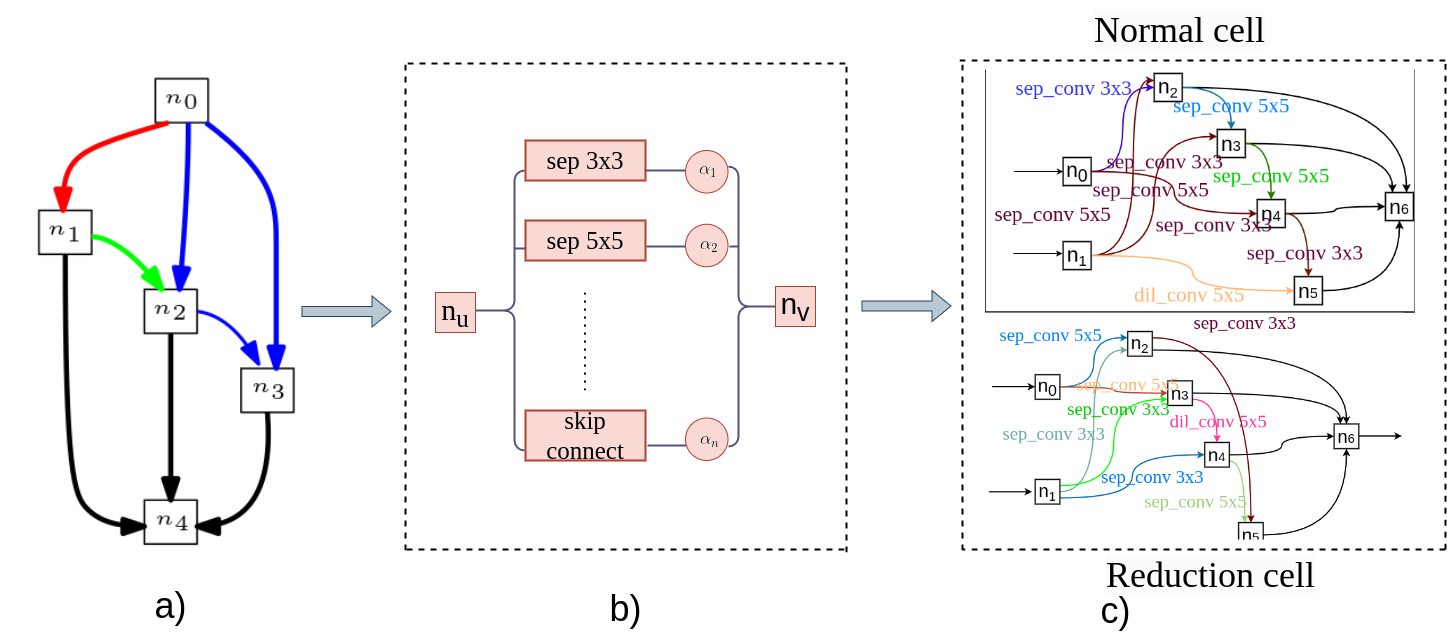}
    \caption{a) illustration of neural architecture search, b)
illustration of search space between node u,v the d) obtained different architecture for normal
and reduction cell}
    \label{fig:block_diagram}\vspace{-0.5cm}
\end{figure*}

\vspace{-0.4cm}
\section{Proposed Method: \textit{NeuralMultiling}}\label{Method}
\vspace{-0.3cm}
The inspiration for a CNN-based search space is the cognizance that architecture engineering with a CNN often pinpoints repeated patterns consisting of convolutional filter banks, nonlinearities, and a judicious selection of connections to accomplish state-of-the-art results. In this section, we introduce the modified NAS, which is automated to find par excellent architectures. First, we introduce the neural cell in Section \ref{neural_cell}, and in Section \ref{cand_ope} we define the candidate operations, in Section \ref{bi_level} and in Section \ref{RDA}, we modify the basic cell architectural parameters of \cite{liu2018darts} in and finally derive the discrete architecture.

\subsection{Neural cell} \label{neural_cell}
A block diagram of the proposed method is shown in Figure. \ref{fig:block_diagram}. First, the search space is composed of nodes, and candidate operations exist between nodes. The operation with the highest weight is selected as the connection operation. After all connected edges and corresponding operations are selected, the final structure is determined. This section describes the process of building this architecture.
To build a scalable architecture, we need 2-types of convolutional cells to deliver 2-main functions when taking a feature map as input:1) Normal cell: A convolutional cell that returns a feature map of the same dimensions and maintains the same number of channels. The normal cell is responsible for capturing features and patterns from the input data and passing them to the next stage of the network. 2) The reduction cell is a convolutional cell that reduces the input feature map by a factor of two using a stride of two for all its operations, in contrast to the normal cell. In addition, this increases the number of channels, resulting in a reduction in the computational cost and complexity of the network. Reduction cells are used to downsample feature maps and compress the information before passing it to the next stage of the network.

A cell is a directed acyclic graph consisting of an ordered sequence of n nodes; in our case, n=7. Each node $n^{(i)}$ is a latent representation (e.g., a feature map on a convolutional neural network) and has directed edges $(u,v)$ associated with some operation $o^{(u,v)}$ that transforms $n^{(i)}$. The structure of each cell was fixed, with each cell having two input nodes, four transitional cells, and one output node.  A neural cell consists of two types of parameters:1) architectural parameters, which specify the structure of a neural cell in terms of the edges (or transformation tensors) connecting the source and target nodes and the operation being performed on them within the cell.  2) Weight parameters: The weight parameters are optimized while keeping the architectural parameters of the cell fixed.

\textbf{Input node:} The input to the $k^{th}$ cell is the output of the last two cells, the first input node $n_0$ is the output of the $(k-1)^{th}$ cell, and the second input node is the output of the $(k-2)^{th}$ cell. Both inputs to the $1^{st}$ cell are the same speech spectrogram. For the $2^{nd}$ cell, the first input $n_0$ is the output from the first cell and the second input is the speech spectrogram.

\textbf{Intermediate node:} All intermediate nodes ($n_2 \rightarrow n_5$) are densely connected, and each intermediate node $n_i$ is computed as the summation of operations based on all its predecessors:
\begin{equation}
    n^v=\sum_{u<v}o^{(u,v)} (n^{(u)})
\end{equation}
Because there were four intermediate nodes in a cell for our experiment, the number of edges within a cell was 14.

\textbf{Output node:} The output from all the intermediate nodes is concatenated to
form the output node. The architecture of a cell with all of its edges is shown in Figure \ref{fig:sample_cell}.

\subsection{Candidate operations} \label{cand_ope}
Each edge, connecting from the input node to the intermediate node and from the
intermediate node to another intermediate node is associated with one of the operations from the set of candidate operations. These operations must be well-defined
and is capable of capturing the variability and generality of data.
The set of candidate operations utilized to derive the architecture are \textit{$'maxpool\:3\times3'$}, \textit{$'avgpool \:3\times3'$}, \textit{$'skip \:connect'$}, \textit{$'sepconv \:3\times 3'$},\textit{$'dilconv\:3\times 3'$},
\textit{$'sepconv \:5\times 5'$} and 
\textit{$'dilconv\:5\times 5'$}



Each of the above operations has a constrained filter size, which we refer to as a channel that can be fine-tuned to obtain the optimal channel size for a given dataset. These operations are common in modern CNN architectures. Finally, each edge was associated with one of these operations. The best combination of operations with these edges is obtained at the end of the search process. It is noteworthy that the operations chosen for each cell were independent of each other. These operations constitute search space ($O$). Our convolutional cell consisted of N=8 neural cells and an initial number of channels C=16; the network was formed by stacking them together. Following previous studies \cite{liu2018darts} reduction cells are located at $\frac{1}{3}$ and $\frac{2}{3}$ positions of the total depth of the network, and the rest are normal cells.

\subsection{Continuous Relaxation over the Cells and Bi-level Optimization}\label{bi_level}
Let $O$ be a set of candidate operations (e.g., convolutions, max pooling, and zero), where each operation refers to a function $o(.)$ to be applied to $n^u$. We use normal and reduction cells to have different architectures, which is contrary to \cite{liu2018darts} and formulate the continuous search space relaxing the categorical choice of operations to be a softmax over all possible operations:
\begin{equation}
   \bar{o}^{(k,u,v)}=\sum_{o\in O}\frac{exp(\alpha_O^{(k,u,v)})}{\sum_{o'\in O} exp(\alpha_{o'}^{(k,u,v)})}o(n)
\end{equation}

\begin{figure*}[!t]
    \centering
    \includegraphics[width=1 \textwidth]{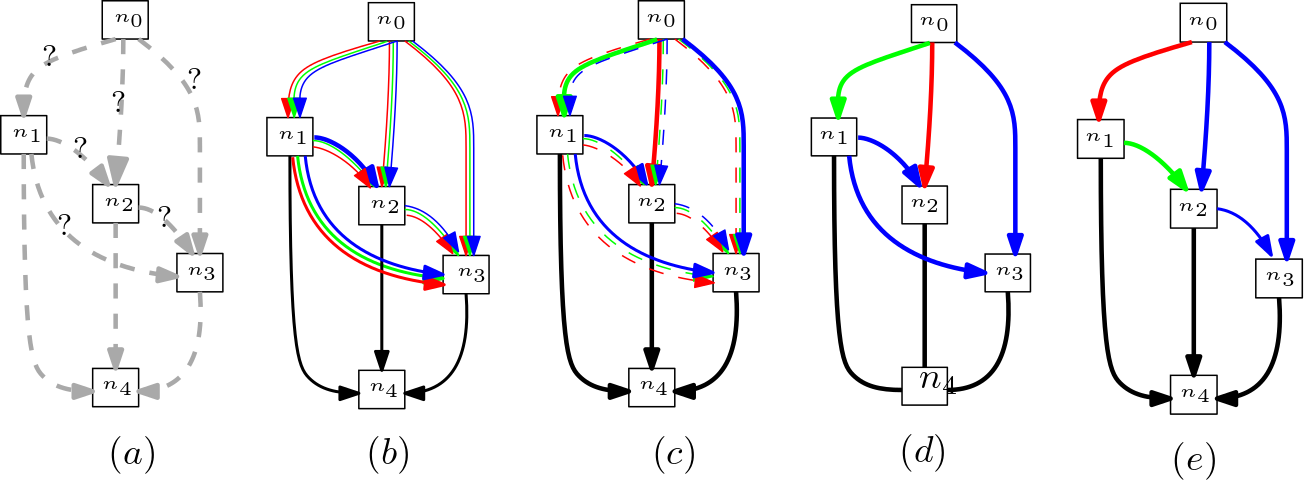}
    \caption{An overview of Continuous relaxation: a) Initial architecture with unknown operations. b) Continuous relaxation of the searched space on each of the edges by setting up candidate operations. c) Two-way optimization of network weights and probabilities of each node. d) $\&$ e) Spawning the final architecture from the learned probabilities for normal cell and reduction cell.}
    \label{fig:bi-level-opt}
\end{figure*}

The goal of the architecture search is then reduced to learning in a continuous variable $\alpha =\{\alpha^{(k,u,v)}\}$ where k is the cell index as illustrated in Figure \ref{fig:bi-level-opt}. After searching among candidate architectures, a discrete architecture is obtained by jointly optimizing the mixing probabilities and network weights by solving the bi-level optimization problem for each normal and reduction cell with the most likely operations, that is, $o^{(k,u,v)}_{normal}=argmax_{o\in O} \hspace{0.2cm}\alpha_o^{(k,u,v)}$ for normal cells and $o^{(k,u,v)}_{reduction}=argmax_{o\in O} \hspace{0.2cm}\alpha_o^{(k,u,v)}$ for the reduction cell, contrary to \cite{liu2018darts}. Subsequently, we aim to jointly learn the candidate architectures and weight parameters.
After soothing out, we aim to jointly learn the candidate architectures and weight parameters.

In each iteration of the Algorithm \ref{searchAlgo}  two steps are being carried out for each cell(k):
\begin{itemize}
    \item \textbf{Weight parameter ($\omega$) update:} During this step, the weight parameters are
optimized while keeping the architectural parameters of the cell fixed.
\item \textbf{Architectural parameter ($\alpha^k$) update:} In this step, the architectural parameters of the cell are updated based on the architectural loss while fixing the
weight parameters.
The update of both the weight and architecture parameters is achieved through
the minimization of the respective cross-entropy loss equation \ref{CEL}. 

\end{itemize}

\begin{algorithm}[!ht]
    
    \caption{\textbf{Search Algorithm:}}\label{searchAlgo}
    \begin{algorithmic}
        \State \textbf{Input: } $\gets$ Training data $\mathcal{D}_{train}$ and validation data  $\mathcal{D}_{val}$ 
        \State \textbf{Output: } Searched Architecture

        \Procedure{Entropy Calculation}{}
            
            
            \begin{equation}
                \label{Ent:eq}
                E = \sum_{k \in C} \sum_{(u,v)} \sum_{o \in \mathcal{O}} \alpha_{kuv}^{o} \log \alpha_{kuv}^{o}
            \end{equation}
            \vspace{-0.5cm}
        \EndProcedure\\
        
        \Procedure{NAS}{}
            \While {\textit{entropy\text{ } decreases}}   \hspace{0.01cm}\Comment{\color{blue} $\backslash*\dots$entropy of \hspace*{4.5cm}the cells have decreased\ $\backslash$} \color{black}
                \For{each cell k}
                    \State \textit{Fix the Architectural Parameters for a cell $(\alpha_{k})$}
                    \State \hspace{0.5cm}$\mathcal{L}_{train} \gets $ training loss $\mathcal{D}_{train}$ 
                    \State \hspace{0.5cm}$\nabla_{\omega} \mathcal{L}_{train} \gets $ gradient on $\mathcal{D}_{train}$ 
                    \State \hspace{0.5cm}Update the weight parameters $(\omega)$
                    \State \textit{Fix the weight parameters$(\omega)$}
                    \State \hspace{0.5cm}$\mathcal{L}_{val} \gets $ Validation loss $\mathcal{D}_{val}$ 
                    \State \hspace{0.5cm}$\nabla_{\alpha_{k}} \mathcal{L}_{val} \gets $ gradient on $\mathcal{D}_{val}$ 
                    \State \hspace{0.5cm}Update the architectural parameters $(\alpha_{k})$
                \EndFor
            \EndWhile
        \EndProcedure 
    \end{algorithmic}
\end{algorithm}

\subsection{Re-defining the architecture parameters of normal and reduction cell}
\label{RDA}

According to \cite{liu2018darts}, normal and reduction cells have the same architecture parameters, that is, $14 (edges) \times 8 (operations)$ for both types of cells. In contrast, we assume that the normal and reduction cells have different architectures by modifying the parameter dimensionality of the normal cell to (number of cells$-2)\times 14(edges)\times8(operations)$ and $2\times 14(edges)\times 8(operations)$ for the reduction cell. By doing so, we increase the search within the specified search space, which better captures the speaker variability across various speakers. More details about the modified architecture is given in the supplementary material.


The main objective of architecture search is to produce an excellent architecture $\alpha^* $ that minimizes the validation loss $L_{val}=(\omega^*,\alpha^*) $ where $\omega^*$ is the weight parameter obtained by minimizing the training loss $\omega^*=argmin_\omega \hspace{0.2cm}L_{train}(\omega,\alpha^*)$. The architecture parameters of the normal and reduction cells are considered to be a 3-D tuple $(x,y,z)$ where $x=$ cell index, $y=$ number of edges, $z=$ number of operations are jointly optimized by passing through the Adam optimizer. The outcome of the optimizer is to produce low entropy, and the cross-entropy losses for $L_{train}$ and $L_{val}$ are described as follows:

\begin{equation}
 L_{CE}=-\sum_{i=1}^{N}t_i log(p_i)
 \label{CEL}
\end{equation}
where $t_i$:ground truth speaker, $N$:Number of speakers, $logp_i$:softmax probability of speaker i.

\subsubsection{Deriving Discrete Architectures}
\label{DDA}
To construct each node in the architecture, we keep the top-2 highest softmax probabilities among all non-zero operations accumulated from previous nodes. The softmax probability of an operation $O$ between the nodes $(u,v)$ is defined as:
\begin{equation}
   p^o_{(k,u,v)}=\frac{exp(\alpha^o_{(k,u,v)})}{\sum_{o'\in O}exp(\alpha^{o'}_{(k,u,v)})}
\end{equation}

\begin{figure*}[!t]
\begin{minipage}[!b]{.50\textwidth}
\centering
\includegraphics[width=1\textwidth]{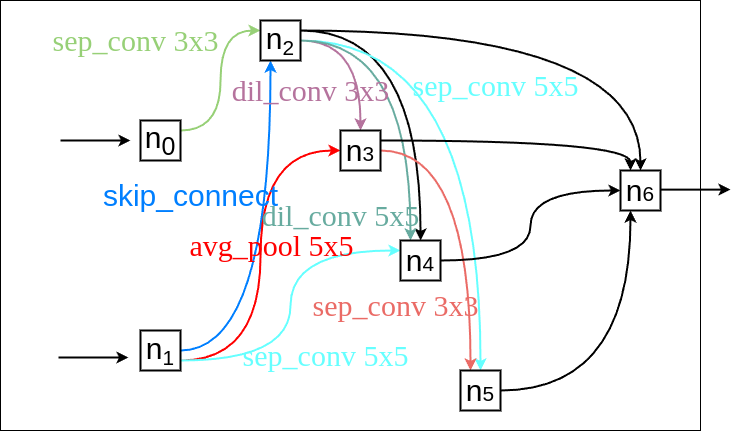}

\caption{Normal cell: Architecture derived from our proposed search algorithm}
\label{fig:Normal_cell}
\end{minipage}
\begin{minipage}[!b]{.50\textwidth}
\centering
\includegraphics[width=1\textwidth]{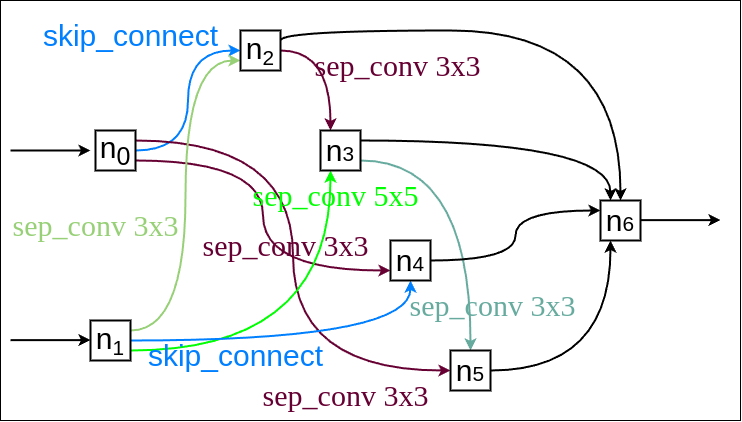}
\caption{Reduction cell: Architecture derived from our proposed search algorithm}
\label{fig:Reduction_cell}
\end{minipage}
\end{figure*}


The visualization of the architectures of the searched normal cell and reduction cell is as shown in Figure \ref{fig:Normal_cell} and \ref{fig:Reduction_cell}. 


\begin{table*}[!th]
\footnotesize
\centering
\caption{Verification performance of the proposed and existing methods for multilingual speaker verification for language agnostic scenario}
\label{tab:EER-LA}
    \begin{tabular}{|ccccccccc}
    \hline
    \multicolumn{9}{c}{\makecell{EER (\%)}} \\ \hline
    \multicolumn{1}{c}{\multirow{3}{*}{Trained on $\downarrow$}} &
      \multicolumn{8}{c}{Tested on} \\ \cline{2-9} 
    \multicolumn{1}{c}{} &
      \multicolumn{4}{c|}{Proposed} &
      \multicolumn{4}{c}{Autospeech \cite{ding2020autospeech}} \\ \cline{2-9} 
    \multicolumn{1}{c}{} &
      \multicolumn{1}{c}{English } &
      \multicolumn{1}{c}{Hindi } &
      \multicolumn{1}{c}{Bengali } &
      \multicolumn{1}{c|}{Parameters} &
      \multicolumn{1}{c}{English } &
      \multicolumn{1}{c}{Hindi } &
      \multicolumn{1}{c}{Bengali} &Parameters
       \\ \hline
    \multicolumn{1}{c}{English} &
      \multicolumn{1}{c}{\textbf{20.99}} &
      \multicolumn{1}{c}{21.33} &
      \multicolumn{1}{c}{23.74} &
      \multicolumn{1}{c|}{362383} &
      \multicolumn{1}{c}{\textbf{27.04}} &
      \multicolumn{1}{c}{25.72} &
      \multicolumn{1}{c}{27.44} &418079
       \\ 
    \multicolumn{1}{c}{Hindi} &
      \multicolumn{1}{c}{22.68} &
      \multicolumn{1}{c}{\textbf{17.73}} &
      \multicolumn{1}{c}{19.75} &
      \multicolumn{1}{c|}{362383} &
      \multicolumn{1}{c}{26.02} &
      \multicolumn{1}{c}{\textbf{22.21}} &
      \multicolumn{1}{c}{24.67} &418079
       \\ 
    \multicolumn{1}{c}{Bengali} &
      \multicolumn{1}{c}{21.95} &
      \multicolumn{1}{c}{19.59} &
      \multicolumn{1}{c}{\textbf{18.95}} &
      \multicolumn{1}{c|}{362383} &
      \multicolumn{1}{c}{25.90} &
      \multicolumn{1}{c}{25.48} &
      \multicolumn{1}{c}{\textbf{23.18}} &418079
       \\ \hline
    \end{tabular}

\end{table*}

\section{Experiments and Results}
\label{sec:exp}
In this section, we discuss the quantitative results of the proposed and existing methods for multilingual speaker verification. First, we present the multilingual dataset employed in this study, followed by the implementation details and discussion of the results. 
\subsection{Multilingual Voice Dataset}
We conducted our experiments on the MAVS database \cite{mandalapu2021multilingual}, which consists of 37,810 utterances in three languages recorded in three different sessions using five different mobile phones: iPhone 6s, iPhone 10s, iPhone 11, Samsung S7, and Samsung S8 spoken by 103 speakers (70 male and 33 female). The three different sessions include session-1 with no noise, session-2 with controlled noise, and session-3 with uncontrolled noise from the natural background. The dataset was divided into $32,250(80\%)$ utterances for training and $5560(20\%)$ utterances for testing in both stages, that is, for the architecture search and training of the searched architecture for multilingual speaker verification.
In the first stage, we search for each cell architecture using our search procedure explained in Algorithm \ref{searchAlgo}. If the entropy calculated using (\ref{Ent:eq}) remains the same for dozens of epochs, we can conclude that the algorithm has converged. In the second stage, the searched architecture was trained from scratch, and its performance on the test set was reported. We used a single-searched architecture for all the  experiments reported in Section \ref{sec:REsults}.

\vspace{-0.2cm}

\subsection{Implementation details}
For each utterance, we excerpted a 257-dimensional spectrogram with a 25ms window and 10ms overlap.
We implemented the proposed architecture search using Pytorch and trained it on a paramshakti supercomputer which has 22 nodes, each of the node has two GPUs of 16 GB named V100 Tesla, and we used one node for training. The NAS search process model described in Section \ref{Method} was trained for 50 epochs, with a batch size of 8. We utilized the Adam optimizer to optimize both the weight $\omega$ and the architecture parameters $\alpha$ by setting the initial learning rate to $10^{-1}$ and the weight decay of the optimizer to $3 \times 10^{-4}$. The entire search process took five days to converge.
In the second stage, the searched architecture was trained from scratch for 200 epochs, with a batch size of 48. The optimizer learning rate was set to $0.15$ the weight decay to $3 \times 10^{-4}$, and the verification process took less than a day.

\begin{table*}[!ht]
\tiny
\caption{Interoperability results in EER(\%) for the proposed method. E-refers to English, H-refers to Hindi, B-refers to Bengali and each entry in table shows the EER.}
\label{tab:intero_ours}
\begin{tabular}{@{}|cccccccccccccllc@{}}

\toprule
\multicolumn{16}{c}{Devices} \\ \midrule
\multicolumn{1}{c}{Trained on $\downarrow$} &
  \multicolumn{3}{c}{iPhone 6s} &
  \multicolumn{3}{c}{iPhone 10s} &
  \multicolumn{3}{c}{iphone 11} &
  \multicolumn{3}{c}{Samsung S7} &
  \multicolumn{3}{c}{Samsung S8} \\ \midrule
\multicolumn{1}{c}{iPhone 6s} &
  \multicolumn{1}{c}{E} &
  \multicolumn{1}{c}{H} &
  \multicolumn{1}{c}{B} &
  \multicolumn{1}{c}{E} &
  \multicolumn{1}{c}{H} &
  \multicolumn{1}{c}{B} &
  \multicolumn{1}{c}{E} &
  \multicolumn{1}{c}{H} &
  \multicolumn{1}{c}{B} &
  \multicolumn{1}{c}{E} &
  \multicolumn{1}{c}{H} &
  \multicolumn{1}{c}{B } &
  \multicolumn{1}{c}{E} &
  \multicolumn{1}{c}{H} & B 
   \\ \midrule
\multicolumn{1}{c}{E} &
  \multicolumn{1}{c}{\textbf{16.71}} &
  \multicolumn{1}{c}{18.96} &
  \multicolumn{1}{c}{22.21} &
  \multicolumn{1}{c}{\textbf{23.25}} &
  \multicolumn{1}{c}{22.35} &
  \multicolumn{1}{c}{23.35} &
  \multicolumn{1}{c}{\textbf{23.45}} &
  \multicolumn{1}{c}{21.26} &
  \multicolumn{1}{c}{21.62} &
  \multicolumn{1}{c}{\textbf{22.45}} &
  \multicolumn{1}{c}{20.16} &
  \multicolumn{1}{c}{22.92} &
  \multicolumn{1}{c}{\textbf{20.19}} &
  \multicolumn{1}{c}{19.60} & 22.10
   \\ 
\multicolumn{1}{c}{H} &
  \multicolumn{1}{c}{17.96} &
  \multicolumn{1}{c}{\textbf{15.72}} &
  \multicolumn{1}{c}{21.63} &
  \multicolumn{1}{c}{26.25} &
  \multicolumn{1}{c}{\textbf{21.13}} &
  \multicolumn{1}{c}{24.45} &
  \multicolumn{1}{c}{25.62} &
  \multicolumn{1}{c}{\textbf{20.18}} &
  \multicolumn{1}{c}{20.62} &
  \multicolumn{1}{c}{23.62} &
  \multicolumn{1}{c}{\textbf{19.62}} &
  \multicolumn{1}{c}{21.68} &
  \multicolumn{1}{c}{21.36} &
  \multicolumn{1}{c}{\textbf{18.72}} & 21.01 
   \\ 
\multicolumn{1}{c}{B} &
  \multicolumn{1}{c}{20.22} &
  \multicolumn{1}{c}{19.72} &
  \multicolumn{1}{c}{\textbf{17.13}} &
  \multicolumn{1}{c}{25.26} &
  \multicolumn{1}{c}{24.45} &
  \multicolumn{1}{c}{\textbf{20.13}} &
  \multicolumn{1}{c}{24.15} &
  \multicolumn{1}{c}{22.62} &
  \multicolumn{1}{c}{\textbf{19.72}} &
  \multicolumn{1}{c}{24.56} &
  \multicolumn{1}{c}{21.27} &
  \multicolumn{1}{c}{\textbf{20.16}} &
  \multicolumn{1}{c}{22.57} &
  \multicolumn{1}{c}{20.61} & \textbf{19.63}
   \\ \midrule
\multicolumn{1}{c}{iphone 10s} & \\ \midrule
\multicolumn{1}{c}{E} &
  \multicolumn{1}{c}{\textbf{26.79}} &
  \multicolumn{1}{c}{24.20} &
  \multicolumn{1}{c}{26.90} &
  \multicolumn{1}{c}{\textbf{19.25}} &
  \multicolumn{1}{c}{15.36} &
  \multicolumn{1}{c}{17.35} &
  \multicolumn{1}{c}{\textbf{19.71}} &
  \multicolumn{1}{c}{17.42} &
  \multicolumn{1}{c}{20.21} &
  \multicolumn{1}{c}{\textbf{21.84}} &
  \multicolumn{1}{c}{26.78} &
  \multicolumn{1}{c}{26.79} &
  \multicolumn{1}{c}{\textbf{21.78}} &
  \multicolumn{1}{c}{22.72} & 23.88  
   \\ 
\multicolumn{1}{c}{H} &
  \multicolumn{1}{c}{27.38} &
  \multicolumn{1}{c}{\textbf{24.97}} &
  \multicolumn{1}{c}{24.85} &
  \multicolumn{1}{c}{19.94} &
  \multicolumn{1}{c}{\textbf{13.99}} &
  \multicolumn{1}{c}{15.45} &
  \multicolumn{1}{c}{18.14} &
  \multicolumn{1}{c}{\textbf{15.61}} &
  \multicolumn{1}{c}{21.42} &
  \multicolumn{1}{c}{21.84} &
  \multicolumn{1}{c}{\textbf{24.12}} &
  \multicolumn{1}{c}{25.31} &
  \multicolumn{1}{c}{25.03} &
  \multicolumn{1}{c}{\textbf{23.84}} & 22.50
   \\ 
\multicolumn{1}{c}{B} &
  \multicolumn{1}{c}{27.85} &
  \multicolumn{1}{c}{24.56} &
  \multicolumn{1}{c}{\textbf{24.27}} &
  \multicolumn{1}{c}{17.24} &
  \multicolumn{1}{c}{15.53} &
  \multicolumn{1}{c}{\textbf{16.74}} &
  \multicolumn{1}{c}{21.18} &
  \multicolumn{1}{c}{16.44} &
  \multicolumn{1}{c}{\textbf{17.52}} &
  \multicolumn{1}{c}{23.95} &
  \multicolumn{1}{c}{22.53} &
  \multicolumn{1}{c}{\textbf{24.54}} &
  \multicolumn{1}{c}{24.60} &
  \multicolumn{1}{c}{27.78} & \textbf{21.94}
   \\ \midrule
 \multicolumn{1}{c}{iPhone 11} & \\ \midrule
 \multicolumn{1}{c}{E} &
  \multicolumn{1}{c}{\textbf{26.53}} &
  \multicolumn{1}{c}{23.24} &
  \multicolumn{1}{c}{26.19} &
  \multicolumn{1}{c}{\textbf{18.49}} &
  \multicolumn{1}{c}{14.91} &
  \multicolumn{1}{c}{17.21} &
  \multicolumn{1}{c}{\textbf{15.55}} &
  \multicolumn{1}{c}{14.96} &
  \multicolumn{1}{c}{17.49} &
  \multicolumn{1}{c}{\textbf{23.40}} &
  \multicolumn{1}{c}{24.77} &
  \multicolumn{1}{c}{23.15} &
  \multicolumn{1}{c}{\textbf{22.61}} &
  \multicolumn{1}{c}{21.88} & 19.03
   \\ 
\multicolumn{1}{c}{H} &
  \multicolumn{1}{c}{26.20} &
  \multicolumn{1}{c}{\textbf{26.14}} &
  \multicolumn{1}{c}{26.22} &
  \multicolumn{1}{c}{19.94} &
  \multicolumn{1}{c}{\textbf{15.95}} &
  \multicolumn{1}{c}{17.03} &
  \multicolumn{1}{c}{14.82} &
  \multicolumn{1}{c}{\textbf{13.77}} &
  \multicolumn{1}{c}{14.81} &
  \multicolumn{1}{c}{21.72} &
  \multicolumn{1}{c}{\textbf{22.71}} &
  \multicolumn{1}{c}{20.87} &
  \multicolumn{1}{c}{20.89} &
  \multicolumn{1}{c}{\textbf{20.49}} & 21.06 
   \\ 
\multicolumn{1}{c}{B} &
  \multicolumn{1}{c}{27.44} &
  \multicolumn{1}{c}{23.42} &
  \multicolumn{1}{c}{\textbf{27.46}} &
  \multicolumn{1}{c}{19.50} &
  \multicolumn{1}{c}{13.43} &
  \multicolumn{1}{c}{\textbf{16.21}} &
  \multicolumn{1}{c}{16.94} &
  \multicolumn{1}{c}{13.24} &
  \multicolumn{1}{c}{\textbf{15.19}} &
  \multicolumn{1}{c}{23.71} &
  \multicolumn{1}{c}{20.79} &
  \multicolumn{1}{c}{\textbf{23.52}} &
  \multicolumn{1}{c}{24.81} &
  \multicolumn{1}{c}{20.34} & \textbf{20.66}
   \\ \midrule
\multicolumn{1}{c}{Samsung S7} & \\ \midrule
\multicolumn{1}{c}{E} &
  \multicolumn{1}{c}{\textbf{25.51}} &
  \multicolumn{1}{c}{26.19} &
  \multicolumn{1}{c}{26.32} &
  \multicolumn{1}{c}{\textbf{24.25}} &
  \multicolumn{1}{c}{25.57} &
  \multicolumn{1}{c}{25.98} &
  \multicolumn{1}{c}{\textbf{24.57}} &
  \multicolumn{1}{c}{25.17} &
  \multicolumn{1}{c}{25.86} &
  \multicolumn{1}{c}{\textbf{18.19}} &
  \multicolumn{1}{c}{19.22} &
  \multicolumn{1}{c}{20.77} &
  \multicolumn{1}{c}{\textbf{18.96}} &
  \multicolumn{1}{c}{19.41} & 20.63
   \\ 
\multicolumn{1}{c}{H} &
  \multicolumn{1}{c}{26.41} &
  \multicolumn{1}{c}{\textbf{24.44}} &
  \multicolumn{1}{c}{25.73} &
  \multicolumn{1}{c}{26.13} &
  \multicolumn{1}{c}{\textbf{25.22}} &
  \multicolumn{1}{c}{26.41} &
  \multicolumn{1}{c}{24.64} &
  \multicolumn{1}{c}{\textbf{22.12}} &
  \multicolumn{1}{c}{24.73} &
  \multicolumn{1}{c}{15.35} &
  \multicolumn{1}{c}{\textbf{15.91}} &
  \multicolumn{1}{c}{15.02} &
  \multicolumn{1}{c}{20.15} &
  \multicolumn{1}{c}{\textbf{19.23}} & 20.11 
   \\ 
\multicolumn{1}{c}{B} &
  \multicolumn{1}{c}{25.23} &
  \multicolumn{1}{c}{25.11} &
  \multicolumn{1}{c}{\textbf{24.92}} &
  \multicolumn{1}{c}{25.52} &
  \multicolumn{1}{c}{24.78} &
  \multicolumn{1}{c}{\textbf{24.63}} &
  \multicolumn{1}{c}{25.12} &
  \multicolumn{1}{c}{24.06} &
  \multicolumn{1}{c}{\textbf{23.21}} &
  \multicolumn{1}{c}{18.16} &
  \multicolumn{1}{c}{17.17} &
  \multicolumn{1}{c}{\textbf{14.63}} &
  \multicolumn{1}{c}{19.42} &
  \multicolumn{1}{c}{19.71} & \textbf{18.63}
   \\ \midrule
   \multicolumn{1}{c}{Samsung S8} &\\ \midrule
   \multicolumn{1}{c}{E} &
  \multicolumn{1}{c}{\textbf{22.12}} &
  \multicolumn{1}{c}{27.18} &
  \multicolumn{1}{c}{21.91} &
  \multicolumn{1}{c}{\textbf{26.54}} &
  \multicolumn{1}{c}{20.57} &
  \multicolumn{1}{c}{20.71} &
  \multicolumn{1}{c}{\textbf{22.88}} &
  \multicolumn{1}{c}{28.59} &
  \multicolumn{1}{c}{24.90} &
  \multicolumn{1}{c}{\textbf{20.06}} &
  \multicolumn{1}{c}{18.19} &
  \multicolumn{1}{c}{17.13} &
  \multicolumn{1}{c}{\textbf{16.94}} &
  \multicolumn{1}{c}{18.60} & 16.09
   \\ 
\multicolumn{1}{c}{H} &
  \multicolumn{1}{c}{20.35} &
  \multicolumn{1}{c}{\textbf{23.61}} &
  \multicolumn{1}{c}{20.26} &
  \multicolumn{1}{c}{24.67} &
  \multicolumn{1}{c}{\textbf{28.41}} &
  \multicolumn{1}{c}{23.06} &
  \multicolumn{1}{c}{20.90} &
  \multicolumn{1}{c}{\textbf{27.58}} &
  \multicolumn{1}{c}{26.94} &
  \multicolumn{1}{c}{17.64} &
  \multicolumn{1}{c}{\textbf{17.14}} &
  \multicolumn{1}{c}{17.35} &
  \multicolumn{1}{c}{18.32} &
  \multicolumn{1}{c}{\textbf{16.19}} & 15.11 
   \\ 
\multicolumn{1}{c}{B} &
  \multicolumn{1}{c}{27.44} &
  \multicolumn{1}{c}{23.42} &
  \multicolumn{1}{c}{\textbf{27.46}} &
  \multicolumn{1}{c}{19.50} &
  \multicolumn{1}{c}{13.43} &
  \multicolumn{1}{c}{\textbf{16.21}} &
  \multicolumn{1}{c}{16.94} &
  \multicolumn{1}{c}{13.24} &
  \multicolumn{1}{c}{\textbf{15.19}} &
  \multicolumn{1}{c}{23.71} &
  \multicolumn{1}{c}{20.79} &
  \multicolumn{1}{c}{\textbf{23.52}} &
  \multicolumn{1}{c}{24.81} &
  \multicolumn{1}{c}{20.34} & \textbf{20.66}
   \\ \bottomrule
  
\end{tabular}%

\end{table*}

\begin{table*}[!ht]
\tiny

\caption{Interoperability results in EER(\%) for Autospeech \cite{ding2020autospeech}. E- refers to English, H-refers to Hindi, B-Bengali language. Each entry in the table indicates EER.}
\label{tab:interop_auto}
\begin{tabular}{@{}|cccccccccccccllc@{}}

\toprule
\multicolumn{16}{c}{Devices} \\ \midrule
\multicolumn{1}{c}{Trained on $\downarrow$} &
  \multicolumn{3}{c}{iPhone 6s} &
  \multicolumn{3}{c}{iPhone 10s} &
  \multicolumn{3}{c}{iphone 11} &
  \multicolumn{3}{c}{Samsung S7} &
  \multicolumn{3}{c}{Samsung S8} \\ \midrule
\multicolumn{1}{c}{iPhone 6s} &
  \multicolumn{1}{c}{E} &
  \multicolumn{1}{c}{H} &
  \multicolumn{1}{c}{B} &
  \multicolumn{1}{c}{E} &
  \multicolumn{1}{c}{H} &
  \multicolumn{1}{c}{B} &
  \multicolumn{1}{c}{E} &
  \multicolumn{1}{c}{H} &
  \multicolumn{1}{c}{B} &
  \multicolumn{1}{c}{E} &
  \multicolumn{1}{c}{H} &
  \multicolumn{1}{c}{B } &
  \multicolumn{1}{c}{E} &
  \multicolumn{1}{c}{H} & B 
   \\ \midrule
\multicolumn{1}{c}{E} &
  \multicolumn{1}{c}{\textbf{18.71}} &
  \multicolumn{1}{c}{20.72} &
  \multicolumn{1}{c}{22.72} &
  \multicolumn{1}{c}{\textbf{25.49}} &
  \multicolumn{1}{c}{26.18} &
  \multicolumn{1}{c}{25.23} &
  \multicolumn{1}{c}{\textbf{25.36}} &
  \multicolumn{1}{c}{26.18} &
  \multicolumn{1}{c}{25.68} &
  \multicolumn{1}{c}{\textbf{24.45}} &
  \multicolumn{1}{c}{21.16} &
  \multicolumn{1}{c}{23.62} &
  \multicolumn{1}{c}{\textbf{22.16}} &
  \multicolumn{1}{c}{21.05} &20.18 
   \\ 
\multicolumn{1}{c}{H} &
  \multicolumn{1}{c}{21.42} &
  \multicolumn{1}{c}{\textbf{18.62}} &
  \multicolumn{1}{c}{23.31} &
  \multicolumn{1}{c}{25.69} &
  \multicolumn{1}{c}{\textbf{23.39}} &
  \multicolumn{1}{c}{24.35} &
  \multicolumn{1}{c}{23.14} &
  \multicolumn{1}{c}{\textbf{22.29}} &
  \multicolumn{1}{c}{21.26} &
  \multicolumn{1}{c}{24.45} &
  \multicolumn{1}{c}{\textbf{21.79}} &
  \multicolumn{1}{c}{22.55} &
  \multicolumn{1}{c}{24.46} &
  \multicolumn{1}{c}{\textbf{20.60}} &23.30 
   \\ 
\multicolumn{1}{c}{B} &
  \multicolumn{1}{c}{20.62} &
  \multicolumn{1}{c}{22.92} &
  \multicolumn{1}{c}{\textbf{17.69}} &
  \multicolumn{1}{c}{24.11} &
  \multicolumn{1}{c}{23.09} &
  \multicolumn{1}{c}{\textbf{22.05}} &
  \multicolumn{1}{c}{21.04} &
  \multicolumn{1}{c}{22.56} &
  \multicolumn{1}{c}{\textbf{20.62}} &
  \multicolumn{1}{c}{23.03} &
  \multicolumn{1}{c}{22.72} &
  \multicolumn{1}{c}{\textbf{22.50}} &
  \multicolumn{1}{c}{22.11} &
  \multicolumn{1}{c}{23.62} & \textbf{21.56}
   \\ \midrule
\multicolumn{1}{c}{iphone 10} & \\ \midrule
\multicolumn{1}{c}{E} &
  \multicolumn{1}{c}{\textbf{24.40}} &
  \multicolumn{1}{c}{25.10} &
  \multicolumn{1}{c}{26.90} &
  \multicolumn{1}{c}{\textbf{20.25}} &
  \multicolumn{1}{c}{19.25} &
  \multicolumn{1}{c}{18.19} &
  \multicolumn{1}{c}{\textbf{20.17}} &
  \multicolumn{1}{c}{18.62} &
  \multicolumn{1}{c}{22.12} &
  \multicolumn{1}{c}{\textbf{23.48}} &
  \multicolumn{1}{c}{25.78} &
  \multicolumn{1}{c}{25.56} &
  \multicolumn{1}{c}{\textbf{22.06}} &
  \multicolumn{1}{c}{23.27} &24.81   
   \\ 
\multicolumn{1}{c}{H} &
  \multicolumn{1}{c}{24.07} &
  \multicolumn{1}{c}{\textbf{23.60}} &
  \multicolumn{1}{c}{23.16} &
  \multicolumn{1}{c}{16.26} &
  \multicolumn{1}{c}{\textbf{15.99}} &
  \multicolumn{1}{c}{17.32} &
  \multicolumn{1}{c}{20.54} &
  \multicolumn{1}{c}{\textbf{16.51}} &
  \multicolumn{1}{c}{22.50} &
  \multicolumn{1}{c}{22.19} &
  \multicolumn{1}{c}{\textbf{26.12}} &
  \multicolumn{1}{c}{24.17} &
  \multicolumn{1}{c}{26.03} &
  \multicolumn{1}{c}{\textbf{25.48}} &23.17 
   \\ 
\multicolumn{1}{c}{B} &
  \multicolumn{1}{c}{23.92} &
  \multicolumn{1}{c}{23.05} &
  \multicolumn{1}{c}{\textbf{22.16}} &
  \multicolumn{1}{c}{19.60} &
  \multicolumn{1}{c}{20.11} &
  \multicolumn{1}{c}{\textbf{18.64}} &
  \multicolumn{1}{c}{22.12} &
  \multicolumn{1}{c}{18.06} &
  \multicolumn{1}{c}{\textbf{18.72}} &
  \multicolumn{1}{c}{24.56} &
  \multicolumn{1}{c}{24.35} &
  \multicolumn{1}{c}{\textbf{25.46}} &
  \multicolumn{1}{c}{25.07} &
  \multicolumn{1}{c}{26.90} & \textbf{23.49}
   \\ \midrule
 \multicolumn{1}{c}{iPhone 11} & \\ \midrule
 \multicolumn{1}{c}{E} &
  \multicolumn{1}{c}{\textbf{24.43}} &
  \multicolumn{1}{c}{25.24} &
  \multicolumn{1}{c}{25.14} &
  \multicolumn{1}{c}{\textbf{20.94}} &
  \multicolumn{1}{c}{16.81} &
  \multicolumn{1}{c}{18.12} &
  \multicolumn{1}{c}{\textbf{17.45}} &
  \multicolumn{1}{c}{15.19} &
  \multicolumn{1}{c}{16.69} &
  \multicolumn{1}{c}{\textbf{24.40}} &
  \multicolumn{1}{c}{25.68} &
  \multicolumn{1}{c}{24.17} &
  \multicolumn{1}{c}{\textbf{24.16}} &
  \multicolumn{1}{c}{22.98} &20.11 
   \\ 
\multicolumn{1}{c}{H} &
  \multicolumn{1}{c}{24.24} &
  \multicolumn{1}{c}{\textbf{24.44}} &
  \multicolumn{1}{c}{25.22} &
  \multicolumn{1}{c}{20.14} &
  \multicolumn{1}{c}{\textbf{17.59}} &
  \multicolumn{1}{c}{16.17} &
  \multicolumn{1}{c}{16.92} &
  \multicolumn{1}{c}{\textbf{15.71}} &
  \multicolumn{1}{c}{15.19} &
  \multicolumn{1}{c}{22.06} &
  \multicolumn{1}{c}{\textbf{23.81}} &
  \multicolumn{1}{c}{21.78} &
  \multicolumn{1}{c}{22.78} &
  \multicolumn{1}{c}{\textbf{22.98}} &22.60  
   \\ 
\multicolumn{1}{c}{B} &
  \multicolumn{1}{c}{25.44} &
  \multicolumn{1}{c}{23.42} &
  \multicolumn{1}{c}{\textbf{24.19}} &
  \multicolumn{1}{c}{21.60} &
  \multicolumn{1}{c}{14.34} &
  \multicolumn{1}{c}{\textbf{18.08}} &
  \multicolumn{1}{c}{18.49} &
  \multicolumn{1}{c}{15.42} &
  \multicolumn{1}{c}{\textbf{17.11}} &
  \multicolumn{1}{c}{24.68} &
  \multicolumn{1}{c}{21.55} &
  \multicolumn{1}{c}{\textbf{23.60}} &
  \multicolumn{1}{c}{25.18} &
  \multicolumn{1}{c}{21.43} & \textbf{22.66}
   \\ \midrule
\multicolumn{1}{c}{Samsung S7} & \\ \midrule
\multicolumn{1}{c}{E} &
  \multicolumn{1}{c}{\textbf{24.41}} &
  \multicolumn{1}{c}{25.91} &
  \multicolumn{1}{c}{25.32} &
  \multicolumn{1}{c}{\textbf{25.52}} &
  \multicolumn{1}{c}{25.17} &
  \multicolumn{1}{c}{25.68} &
  \multicolumn{1}{c}{\textbf{25.75}} &
  \multicolumn{1}{c}{24.17} &
  \multicolumn{1}{c}{26.68} &
  \multicolumn{1}{c}{\textbf{20.91}} &
  \multicolumn{1}{c}{20.23} &
  \multicolumn{1}{c}{21.76} &
  \multicolumn{1}{c}{\textbf{20.68}} &
  \multicolumn{1}{c}{20.14} &22.40 
   \\ 
\multicolumn{1}{c}{H} &
  \multicolumn{1}{c}{24.61} &
  \multicolumn{1}{c}{\textbf{25.44}} &
  \multicolumn{1}{c}{25.63} &
  \multicolumn{1}{c}{26.31} &
  \multicolumn{1}{c}{\textbf{25.24}} &
  \multicolumn{1}{c}{27.72} &
  \multicolumn{1}{c}{25.46} &
  \multicolumn{1}{c}{\textbf{23.21}} &
  \multicolumn{1}{c}{24.37} &
  \multicolumn{1}{c}{16.53} &
  \multicolumn{1}{c}{\textbf{16.76}} &
  \multicolumn{1}{c}{17.20} &
  \multicolumn{1}{c}{22.61} &
  \multicolumn{1}{c}{\textbf{21.64}} &21.07  
   \\ 
\multicolumn{1}{c}{B} &
  \multicolumn{1}{c}{25.13} &
  \multicolumn{1}{c}{24.11} &
  \multicolumn{1}{c}{\textbf{24.62}} &
  \multicolumn{1}{c}{24.22} &
  \multicolumn{1}{c}{23.87} &
  \multicolumn{1}{c}{\textbf{24.36}} &
  \multicolumn{1}{c}{25.42} &
  \multicolumn{1}{c}{25.60} &
  \multicolumn{1}{c}{\textbf{24.21}} &
  \multicolumn{1}{c}{20.80} &
  \multicolumn{1}{c}{19.76} &
  \multicolumn{1}{c}{\textbf{15.40}} &
  \multicolumn{1}{c}{21.36} &
  \multicolumn{1}{c}{22.61} & \textbf{20.11}
   \\ \midrule
   \multicolumn{1}{c}{Samsung S8} &\\ \midrule
   \multicolumn{1}{c}{E} &
  \multicolumn{1}{c}{\textbf{23.21}} &
  \multicolumn{1}{c}{27.81} &
  \multicolumn{1}{c}{22.60} &
  \multicolumn{1}{c}{\textbf{27.13}} &
  \multicolumn{1}{c}{22.62} &
  \multicolumn{1}{c}{21.18} &
  \multicolumn{1}{c}{\textbf{23.66}} &
  \multicolumn{1}{c}{26.95} &
  \multicolumn{1}{c}{25.68} &
  \multicolumn{1}{c}{\textbf{18.16}} &
  \multicolumn{1}{c}{20.22} &
  \multicolumn{1}{c}{19.46} &
  \multicolumn{1}{c}{\textbf{18.19}} &
  \multicolumn{1}{c}{20.60} &17.11 
   \\ 
\multicolumn{1}{c}{H} &
  \multicolumn{1}{c}{21.55} &
  \multicolumn{1}{c}{\textbf{24.09}} &
  \multicolumn{1}{c}{21.08} &
  \multicolumn{1}{c}{25.69} &
  \multicolumn{1}{c}{\textbf{27.51}} &
  \multicolumn{1}{c}{24.60} &
  \multicolumn{1}{c}{21.62} &
  \multicolumn{1}{c}{\textbf{26.65}} &
  \multicolumn{1}{c}{27.18} &
  \multicolumn{1}{c}{18.56} &
  \multicolumn{1}{c}{\textbf{19.14}} &
  \multicolumn{1}{c}{20.46} &
  \multicolumn{1}{c}{20.22} &
  \multicolumn{1}{c}{\textbf{18.22}} &16.44  
   \\ 
\multicolumn{1}{c}{B} &
  \multicolumn{1}{c}{26.44} &
  \multicolumn{1}{c}{25.32} &
  \multicolumn{1}{c}{\textbf{26.64}} &
  \multicolumn{1}{c}{20.66} &
  \multicolumn{1}{c}{15.56} &
  \multicolumn{1}{c}{\textbf{15.76}} &
  \multicolumn{1}{c}{18.11} &
  \multicolumn{1}{c}{15.98} &
  \multicolumn{1}{c}{\textbf{16.12}} &
  \multicolumn{1}{c}{22.66} &
  \multicolumn{1}{c}{25.25} &
  \multicolumn{1}{c}{\textbf{25.18}} &
  \multicolumn{1}{c}{25.62} &
  \multicolumn{1}{c}{21.24} & \textbf{21.88}
   \\ \bottomrule

\end{tabular}%

\end{table*}



\vspace{-0.4cm}
\subsection{Results and Discussion}
\label{sec:REsults}
In this section, we present quantitative results of the proposed method for multilingual speaker verification. The performance of the proposed method was compared with that of Autospeech \cite{ding2020autospeech}, which is based on neural searching. Autospeech was trained using the MAVS \cite{mandalapu2021multilingual} dataset under similar training conditions as described in Autospeech for a fair comparison. There was another model based on Bi-LSTMs \cite{chojnacka2021speakerstew} where they presented a lighweight speaker verification models operated on 46 languages. But to do comparison neither the code nor the dataset is available in open source. We present two different experiments: (1) Language agnostic, in which the speaker is enroled with one language and probed with another language. (2) Interoperability across smartphones (or devices) and languages in which the speaker is enrolled with one device and one language and probed with other devices and languages. The performance of the proposed method is presented using an Equal Error Rate (EER(\%)), which corresponds to the False Match Rate (FMR), and is equal to the False Non-Match Rate (FNMR). 
Table \ref{tab:EER-LA} shows the quantitative performance of the proposed and existing methods in language-agnostic experiments. Here we consider language based speech files from all devices and perform cross language testing. Based on the results in Table \ref{tab:EER-LA}, the following can be observed.

\begin{itemize}
\item The verification performance of the proposed and the existing method indicates the improved performance when trained and tested with the same language.  The best performance was observed when trained and tested using Hindi. 
\item  The verification performance degradation is noted with the proposed and the existing method during the cross language test. It can also be observed that the training language can influence the verification performance of the proposed and existing methods. For example, training with the English language indicated less verification performance degradation when tested with other languages such as Hindi and Bengali. Furthermore, it is interesting to note that the cross-language verification performances of the proposed and existing methods are less degraded between Hindi and Bengali. This can be attributed to similarities in language characteristics.
\item The proposed method indicates the best performance compared to the existing method on cross and same language experiments. 

\begin{figure*}[!h]
    \centering
    \includegraphics[width=0.5\textwidth]{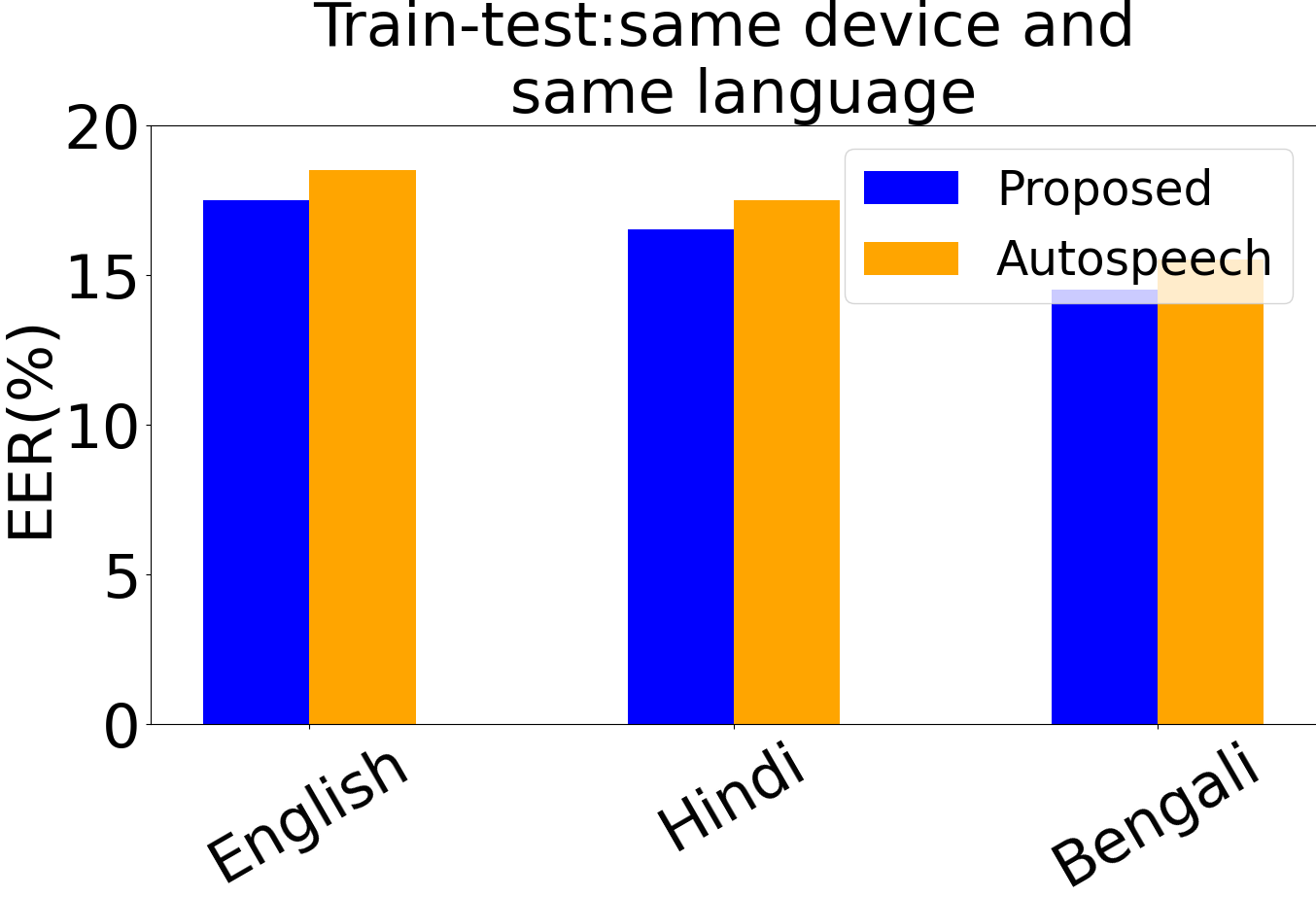}
    \caption{Case-1: Histogram for same device and same language}
    \label{fig:case-1}
\end{figure*}

\item Achieving a language agnostic condition in multilingual speaker verification involves addressing challenges such as variations in phonetic structures, acoustic characteristics and linguistic patterns across different languages. The Autospeech \cite{ding2020autospeech} which follows same architecture for normal cell and reduction cell do not capture the above said characteristics hence a degradation of EER is observed whereas for our proposed method the EER is reduced when cross-language testing is performed. This accounts for robust and a generalizable muiltilingual verification model that can adapt to inherent diversity in languages while maintaining lesser number of parameters.

\begin{figure*}[!h]
    \centering
    \includegraphics[width=1.20\textwidth]{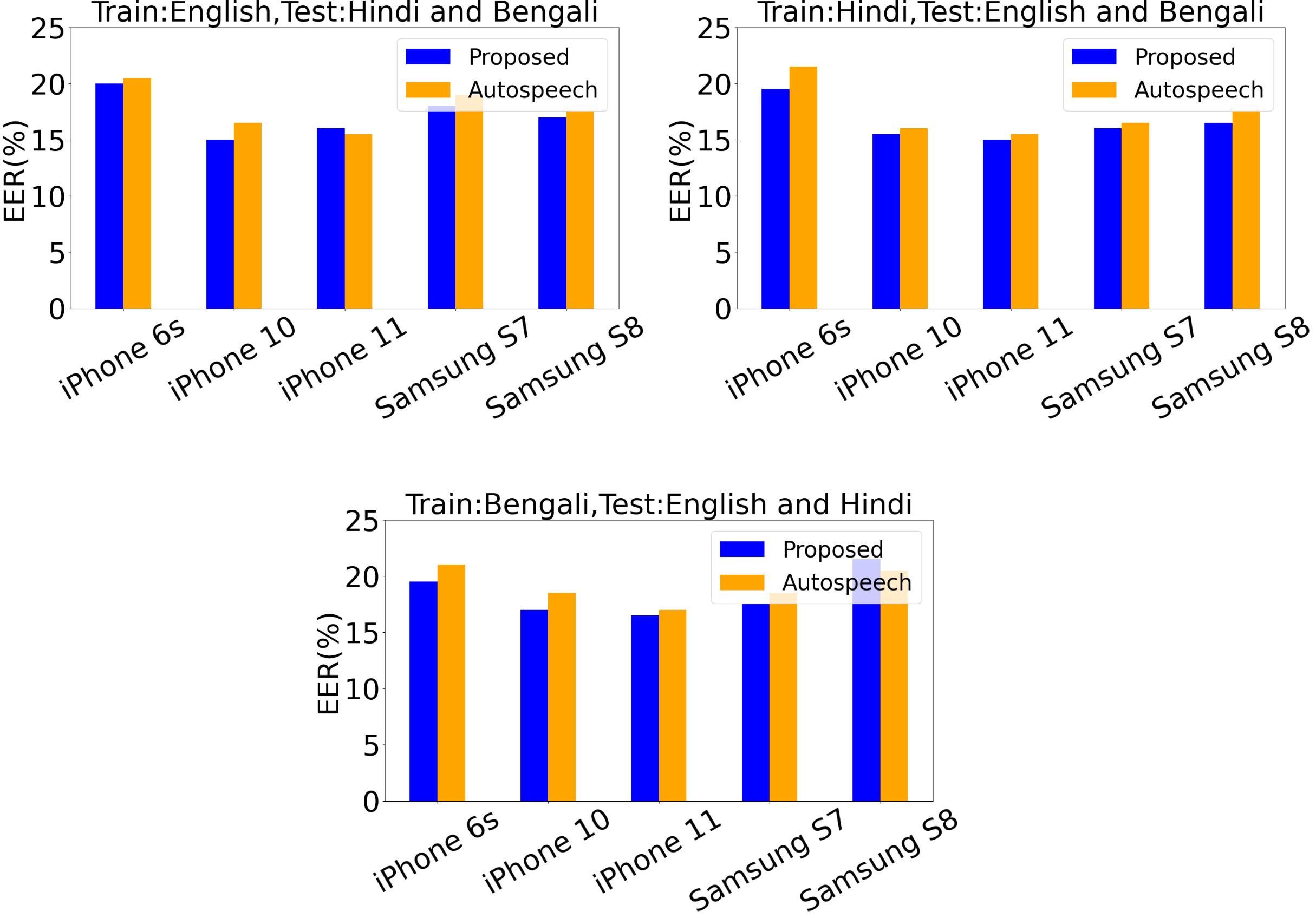}
    \caption{Case-2: Histogram for cross language and same device}
    \label{fig:case-2}
\end{figure*}
\item The proposed method also results in the less number of parameters (362k) and model size of \textit{6.28Mb} compared to the existing method with (418k) parameters with model size of \textit{8.28Mb}. Thus, the proposed method not only outperforms the existing method but also results in a lightweight model suitable for deployment in a smartphone environment. 
\end{itemize}

\begin{figure*}[!th]
    \centering
    \includegraphics[width=1.20\textwidth]{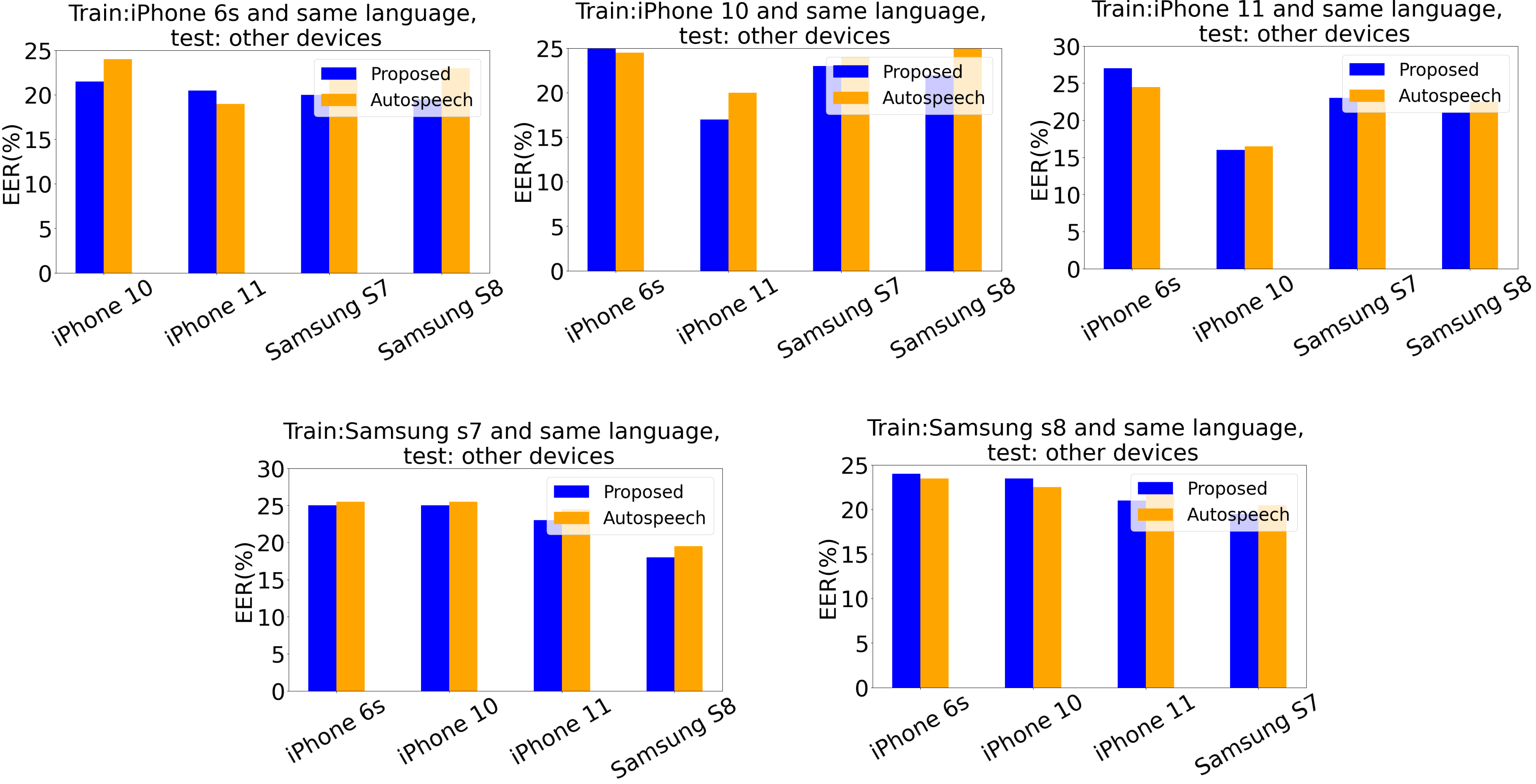}
    \caption{Case 3: Histogram for same language and cross device}
    \label{fig:case-3}
\end{figure*}

Tables \ref{tab:intero_ours} and \ref{tab:interop_auto} show the quantitative performance of the proposed method and Autospeech for interoperability across devices and languages, respectively. The interoperability experimental results were interpreted based on the four cases discussed below.

\begin{itemize}
\item \textbf{Case-I: Same device and same language:} Here, we analyze the verification performance of the proposed method when same language is trained and tested by the same device. This analysis provides insight into the verification performed on independent languages.  Figure \ref{fig:case-1} shows the average EER(\%) with respect to different devices, which is independent of language. As shown in Figure \ref{fig:case-1}, Hindi had the lowest EER(\%), and English had the highest EER(\%). The best performance with the Hindi language can be attributed to the fact that the majority of speakers in the MAVS dataset were native Hindi speakers.

\begin{figure*}[!h]
    \centering
    \includegraphics[width=1.20\textwidth]{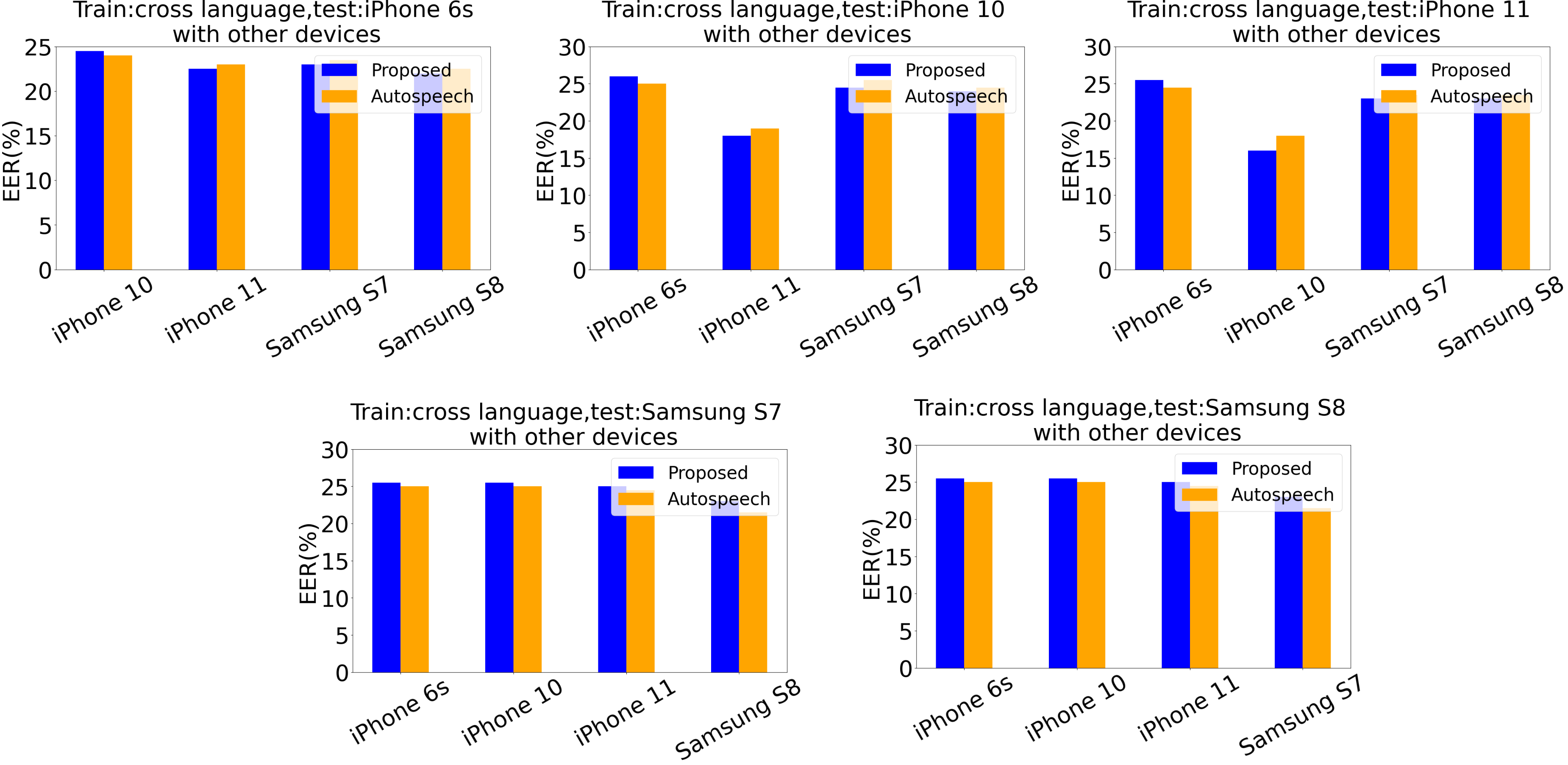}
    \caption{Case-4: Histogram for cross language and cross device}
    \label{fig:case-4}
\end{figure*}

\item   \textbf{Case II: Cross language and same device:} Here, we analyze the verification performance of the proposed method when individual devices are trained in one language and tested in another language. Figure \ref{fig:case-2} shows the verification performance of the proposed method with cross-language and the same device scenario. This experiment allowed us to analyze the interoperability of language across devices.  The obtained results indicate that (a) the verification performance is influenced by the language of the individual devices. (b) The iPhone6S has the highest EER(\%) across all three languages. (3) The iPhone11 indicated the best performance across all three languages and thus emerged as the best language-agnostic device with the proposed method. 

\item \textbf{Case III: Same language and cross device:}  Here, we analyze the verification performance of the proposed method when the same language is used for training and testing, while cross devices are used for verification. Figure \ref{fig:case-3} shows the verification performance of the proposed method when the voice data (irrespective of the language) from one type of device are used for training, and testing is performed using the voice data (same language as that of training) collected from another device. This experiment allowed the interoperability of the devices to be analyzed when the same language was used for training and testing. The obtained results indicated the influence of the device data on the verification performance. In some cases, the interoperability of the devices  indicated improved performance (for example, when trained with the iPhone6S and tested with Samsung S8). However, the verification performance across other devices was less influenced, particularly when Samsung devices were used. 





\begin{table}[!th]
\centering
\scriptsize
\caption{The ablation study of the proposed method and Autospeech \cite{ding2020autospeech} with varying number of nodes and channels such that N indicated number of nodes and C indicates number of channels.}
\label{tab:ablation}
\begin{tabular}{|ccc}
\hline
\multicolumn{1}{c|}{\textbf{Method}} & \multicolumn{1}{c|}{\textbf{No of Parameters}} & \textbf{\makecell{Search cost \\ No of GPU days} } \\ \hline
\multicolumn{3}{c}{Autospeech \cite{ding2020autospeech}}                            \\ \hline
\multicolumn{1}{c|}{N=8,C=16} & \multicolumn{1}{c|}{418k} & 7 \\ \hline
\multicolumn{1}{c|}{N=8, C=64} & \multicolumn{1}{c|}{617k} & 9 \\ \hline
\multicolumn{1}{c|}{N=30, C=64} & \multicolumn{1}{c|}{986k} & 10 \\ \hline
\multicolumn{1}{c|}{N=8, C=128} & \multicolumn{1}{c|}{1160k} &11  \\ \hline
\multicolumn{3}{c}{Proposed}                            \\ \hline
\multicolumn{1}{c|}{N=8,C=16} & \multicolumn{1}{c|}{362k} & 5 \\ \hline
\multicolumn{1}{c|}{N=8, C=64} & \multicolumn{1}{c|}{418k} &  6\\ \hline
\multicolumn{1}{c|}{N=30, C=64} & \multicolumn{1}{c|}{568k} &  8\\ \hline
\multicolumn{1}{c|}{N=8, C=128} & \multicolumn{1}{c|}{625k} & 8.2 \\ \hline
\end{tabular}
\end{table}

\item \textbf{Case IV: Cross language and cross device:} Here, we analyze the verification performance of the proposed method with cross language (training and testing with different languages) and cross device (enrolment using one device and probe with other devices). Figure \ref{fig:case-4} shows the verification performance of the proposed method in cross-language and cross-device scenarios. 
This experiment allowed us to analyze the performance of the proposed system with interoperability for both language and device. Note that (a) the interoperability of the devices indicates higher error rates with cross-language. (b) Verification performance degrades across all devices.
For all four cases, our proposed method outperforms Autospeech \cite{ding2020autospeech} because our proposed model can capture the speaker characteristics better than Autospeech \cite{ding2020autospeech}.

\end{itemize}

\vspace{-0.6cm}
\section{Ablation studies}
To verify the effectiveness of the proposed method, we use original Autospeech to search on our MAVS dataset. It is obvious that the search cost is greatly reduced through our proposed method. We also varied the number of nodes and number channels during the search process and for each model our proposed method outperforms Autospeech both in terms of number of parameters and also in terms of search cost. The experimental results are as shown in Table \ref{tab:ablation}.

\vspace{-0.3cm}
\section{Conclusions}
\label{concl}
In this study, we propose an automatic approach to determine the optimal CNN architecture for multilingual speaker verification. We modified the baseline approach by introducing different architectures for normal and reduction cells. With this modification, we searched for an excellent CNN architecture for neural cells with different edge operations. Subsequently, with the derived architecture we conducted two different experiments: language-agnostic conditions across various smartphone devices, and interoperability by building language models across different devices and languages on MAVS database. For the language-agnostic condition, our proposed method outperformed the baseline model while maintaining lower model complexity. For interoperability, the proposed model also yields better performance when the trained and test mobile phones are from the same manufacturer; however, for cross devices, a slightly higher EER is observed. Overall, we obtained an automatic architecture that is lightweight and performs better than the baseline model, which can be further deployed into mobile devices for multilingual speaker verification.

\bibliographystyle{elsarticle-num}
\bibliography{mybibfile}





\end{document}


\title{(Supplementary) NeuralMultiling: A Novel Neural Architecture Search for Smartphone based Multilingual Speaker Verification}
%
%
\author{Aravinda Reddy PN\inst{1}\orcidID{0000-0002-1342-924x} \and
Raghavendra Ramachandra\inst{2}\orcidID{0000-0003-0484-3956} \and K.Sreenivasa Rao \inst{1}\orcidID{0000--0001-6112-6887} \and Pabitra Mitra\inst{1}\orcidID{0000--0002-1908-9813}   }
\authorrunning{Aravinda et al.}
%
\institute{Indian Institute of Technology Kharagpur, India\\
\email{aravindareddy.27@iitkgp.ac.in}\\
\and
Norwegian University of Science and Technology (NTNU), Norway.}
\maketitle              
\section{Modified architecture}
\subsection{Continous Relaxation Over the cells and Bi-Level Optimization }

Our modified architecture use different architecture for normal cell and reduction cell contrary to Autospeech. So the goal of the architecture search then reduces to learning the set of continuous variables for both normal cell and reduction cell which is contrary to Autospeech.

\subsection{Re-defining the architecture parameters of normal and reduction cell}
\begin{figure*}[!h]
    \centering
    \includegraphics[width=1\textwidth]{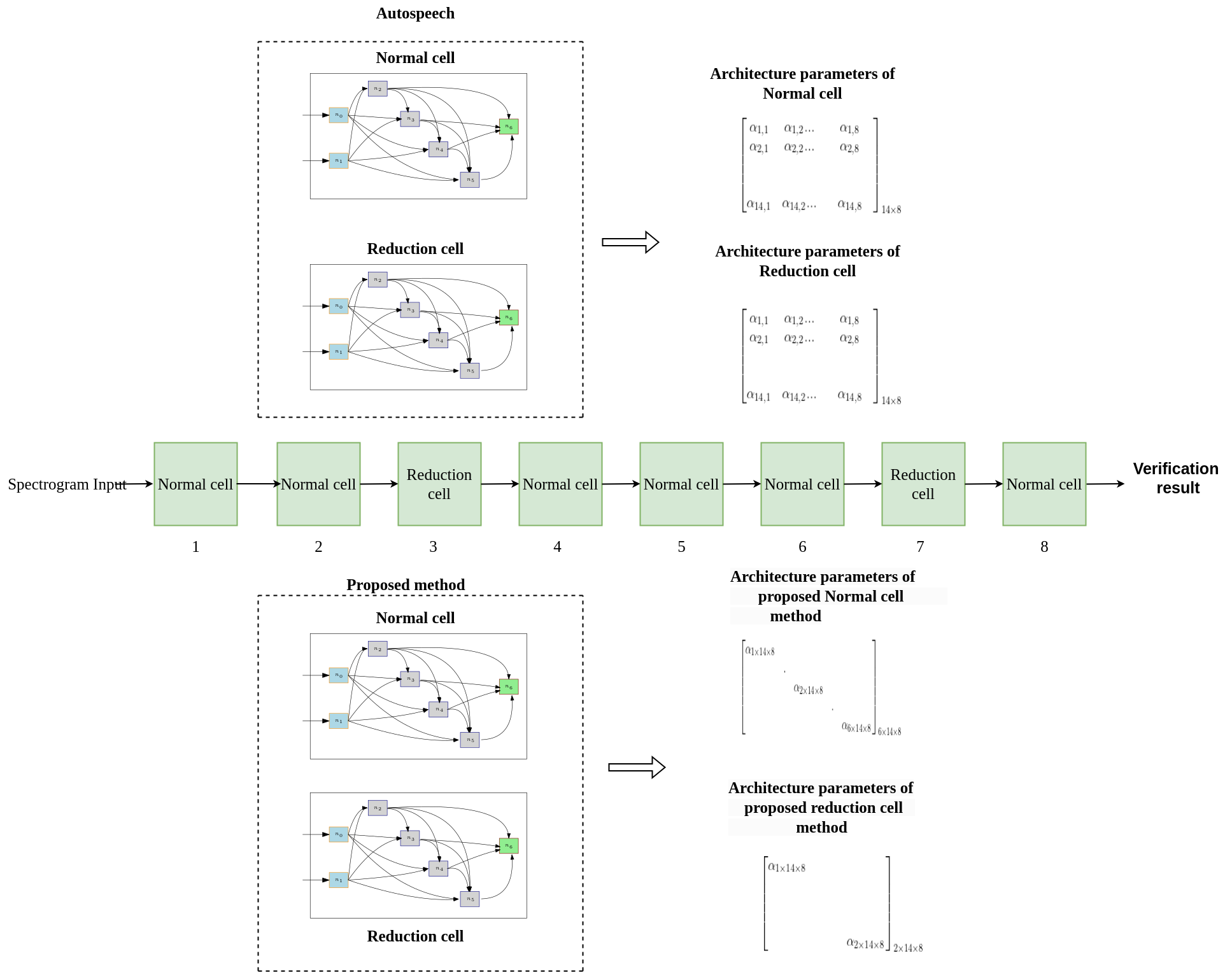}
    \caption{An overview of Autospeech and proposed method}
    \label{fig:proposed_method}
\end{figure*}

\begin{figure*}[!h]
    \centering
    \includegraphics[width=1\textwidth]{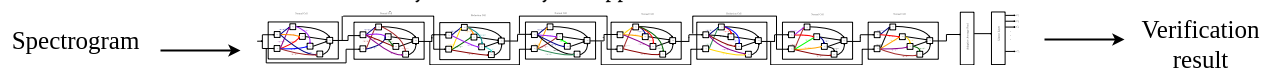}
    \caption{Final derieved CNN architecture with fully connected layer appended }
    \label{fig:fianl_arc}
\end{figure*}

The architecture parameters of normal cell and reduction remains same throughout the search process, that is $14 (edges)\times 8 (operations)$ for both types of cells. We modify the architecture parameters of normal and reduction cell to have different architectures that is normal cell has $6\times14\times8$ and reduction cell to have $2\times14\times 8$ parameters. The modified and existing architecture parameters are as shown
in Figure \ref{fig:proposed_method}. By modifying the architectural parameters we are allowing the model to learn speaker specific characteristics more robustly than the existing method. The final architecture is depicted in Figure \ref{fig:fianl_arc} with fully connected layer appended at the last.

